 \documentstyle[preprint,aps,prb,amsfonts]{revtex}

\begin{document}

\draft
\title{Orthogonal  localized wave functions
 of an electron in a magnetic field}
\author{E. I. Rashba$^{1, 2}$, L. E. Zhukov$^1$, and A. L. Efros$^1$}
\address{
 $^1$Department of Physics, University of Utah, Salt Lake City, UT 84112\\
and~ $^2$L. D. Landau Institute for Theoretical Physics, Moscow 117940}

\date{\today}

\maketitle


\begin{abstract}
  We prove the existence of a set of two-scale magnetic Wannier orbitals,
 $w_{mn}({\bf r})$,
 in the infinite plane. The quantum numbers of these states
 are the positions $(m, n)$
 of their centers which form a von Neumann lattice.
 Function $w_{00}({\bf r})$ localized at the origin has a nearly 
 Gaussian shape of 
 $\exp(-r^{2}/4l^{2})/\sqrt{2\pi}$ for $r \alt \sqrt{2\pi} l$,
 where $l$ is the magnetic length. This region makes a dominating contribution
 to the normalization integral.  Outside this region function,
 $w_{00}({\bf r})$ is small, oscillates, and falls off with
 the Thouless critical exponent for magnetic orbitals, $r^{-2}$. These
 functions form a convenient basis  for many electron problems.

\end{abstract}


\section{Introduction}
\label{sec:introduction}

 Computational methods for interacting electrons in a strong magnetic
 field have been developing rapidly in  the last decade.\cite{CP} They
 employ different geometries which suggest different sets of single-electron
 wave functions and quantum numbers of many-electron states. 
 The toroidal geometry was advanced by Yoshioka, Halperin, and Lee.\cite{YHL}
 The symmetry classification of the many-electron functions in
 this geometry was proposed by Haldane.\cite{Haldane} Afterwards most of
 the finite-size computations were performed in the framework of the spherical
 geometry  proposed and developed by Haldane and Rezayi.\cite{HR}
 
 Landau functions, eigenfunctions of  angular momentum, and localized
 functions called coherent states\cite{MZ,F95,HIR} are usually used in 
 the plane geometry.
  Localized functions have obvious  advantages as applied to
 the Wigner crystal theory and also to some systems with strongly
 inhomogeneous potential distribution. We expect that these functions
 may also be effective  as the basis functions for
 numerical diagonalization. Since 
 the matrix size increases exponentially with the number of particles,
 $N$, this number is strongly  restricted by computational facilities,
 e.g., by $N\leq 10$ for the
 filling factor $\nu =1/3$. Therefore, it seems tempting to develop 
 some procedure for a strong reduction of  the dimension
 of the Hilbert space by eliminating functions
 which make only a small contribution to the low-energy states. We believe
 that the prospects for such an approach are promising since the
 correlation energy, $\varepsilon_{\rm corr}$, which can be evaluated
 by the magnitude of the Laughlin gap or by the energy difference between
 the liquid and solid phases, is small compared with the characteristic
 Coulomb energy, $\varepsilon_{C} = e^{2}/{\epsilon l}$,
 where $l=(c{\hbar}/eB)^{1/2}$
 is the magnetic length. The inequality
 $\varepsilon_{\rm corr} \ll \varepsilon_{C}$ implies that the subspace
 chosen for diagonalization can be restricted by
 the electron configurations with low Hartree-Fock energies. Such an
 approach has been recently developed and successfully
 applied to electrons on a lattice.\cite{efr}

 As a first step in developing a low-energy states selecting procedure,
 one should construct  a basis of localized, particle-like 
 functions with a characteristic size about $l$.
 However, some general requirements exclude the existence of a complete
 system of single-electron states which are both exponentially localized
 and linearly independent. This point should be explained in more detail.
 
 The coherent states\cite{CSR} on a von Neumann lattice are the
 most localized eigenfunctions for an electron in a magnetic field.
 For such a lattice, with a single flux quantum per a unit cell,
 the area of a unit cell equals $2\pi l^2$.
 The set of coherent states, $c_{mn}({\bf r})$, of the lowest Landau level (LLL)
 can be obtained by the magnetic translations of  the function
 $c_{00}(r) = \exp( - r^{2}/4l^{2})/\sqrt{2\pi}$ from the origin  to
 all lattice sites, $(m, n)$. The set $c_{mn}({\bf r})$ is complete,
 i.e., an arbitrary function belonging to the LLL can be expanded
 in functions $c_{mn}({\bf r})$.
 This statement is physically appealing and is supported by a rigorous
 mathematical proof.\cite{Per,BBGK} The functions $c_{mn}({\bf r})$
 are obviously non-orthogonal. 
 A straightforward  way to orthogonalize such a set is to  transform it  into 
 the Bloch representation, and then  go back to the site representation.
 Wannier applied this
 procedure  to a one-dimensional chain
 of Gaussians.\cite{Wann} 
 If the functions $c_{mn}({\bf r})$ were linearly independent, this orthogonalization
procedure would result in a set of exponentially localized functions.
 However, Perelomov\cite{Per} has established a non-trivial fact
 that the set $c_{mn}({\bf r})$  is actually overcomplete by 
 exactly one function and
 presented the explicit form of the linear equation relating these functions; see Eq.~(\ref{eqOC}) below.
 The overcompleteness of the set $c_{mn}({\bf r})$ imposes hard
 restrictions on the localization of Wannier functions. Zak and
 collaborators\cite{DaZ,Zak} have shown that the exponential localization
 and the orthogonality of magnetic orbitals are incompatible. Thouless\cite{Thou}
  proved even a stronger  result by relating the localization
 of magnetic orbitals to the existence of the Hall current. He
 showed that in  systems supporting a Hall current the orbitals should
 fall off with distance no faster than by the inverse-square law, hence,
 $r^{-2}$ is a critical exponent.

 Having in mind these restrictions on the falloff of magnetic orbitals,
 one can question whether particle-like orbitals suitable for
 restricted-basis finite-size calculations do exist. It is the main result
 of this paper that a set of two-scale orbitals $w_{mn}({\bf r})$,
 which satisfy the above requirements exists.\cite{ZRE}
 In the small $\bf r$ region,
 $r\alt \sqrt{2\pi}l$, the orbital $w_{00}({\bf r})$
 is very close to the Gaussian
 $c_{00}(r)$, and this region contributes about 95\% to the
 normalization integral. For $r\agt \sqrt{2\pi}l$, the function 
 $w_{00}({\bf r})$ falls off as $r^{-2}$ in exact correspondence
 with the Thouless criterion and shows an oscillatory behavior which
ensures orthogonality to the different orbitals $w_{mn}({\bf r})$.

 The term Wannier functions is usually applied only to the orbitals
 which fall off exponentially. Since the functions $w_{mn}({\bf r})$
 are normalizable and have a well localized Gaussian core, we shall
 term them as  Wannier functions in what follows.

 The paper is organized as follows. In Sec.~\ref{sec:equations}
 general equations are presented and the behavior of magnetic
 Bloch functions near the singular point in the momentum plane is studied.
 It is of importance for establishing the completeness of the sets of
 Bloch and Wannier functions. The basic results are discussed in
 Sec.~\ref{sec:plane} where two-scale Wannier functions $w_{mn}({\bf r})$
 of the infinite plane are derived and studied analytically and
 numerically. In Sec.~\ref{sec:toroidal} some  results for
 Wannier functions on the finite plaquettes are presented. 
The results are summarized in Conclusion.

\section{General equations}
\label{sec:equations}

  Let us choose a rectangular unit cell with sides $\bf a$ and $\bf b$
 such that $ab = 2\pi$ and a normalization plaquette  with  sides
 $L_{x} = \alpha a$ and $L_{y} = \beta b$, where  $\alpha$ and $\beta$
 are integers. Here and below $l=1$. 
 It is convenient to define the function $c_{mn}({\bf r})$
 centered at a site $(m,n)$  as
\begin{equation}
c_{mn}({\bf r}) = T_{m{\bf a}} T_{n{\bf b}}~ c_{00}(r)~,~~
 c_{00}(r) = \exp(-r^{2}/4)/\sqrt{2\pi}~,
\label{eq0.1}
\end{equation}
 where $T_{m{\bf a}}$ and $T_{n{\bf b}}$ are operators of 
 magnetic translations\cite{BrZak} along $\bf a$ and $\bf b$ axes, and
 $c_{mn}({\bf r})$ are the eigenfunctions of the Schroedinger equation of an electron in a
magnetic field.  The
 axial gauge ${\bf A}({\bf r}) = {1\over 2}~{\hat {\bf z}}\times {\bf r}$ is used.
 The general definition of the operator $T_{\bf R}$ for a particle with
 a charge $e>0$ is:
\begin{equation}
 T_{\bf R}\psi({\bf r)} = \exp\{-{i\over 2}~{\bf R}\cdot{\bf A}({\bf R})\}~
 \exp\{i{\bf r}\cdot {\bf A}({\bf R})\}~ \psi({\bf r}-{\bf R})~.
\label{eq0.2}
\end{equation}
 Therefore, the explicit form of the function $c_{mn}({\bf r})$ is:
\begin{equation}
  c_{mn}({\bf r}) = (-)^{mn}
  ~ \exp\{-{1\over 4} ({\bf r} - {\bf R}_{mn})^{2}
 +~ {i\over 2}~\hat{{\bf z}}\cdot ({\bf R}_{mn}\times {\bf r}) \}
 /\sqrt{2\pi}~,
\label{eq1}
\end{equation}
 where ${\bf R}_{mn} = m{\bf a} + n{\bf b}$. 

 The set of functions $ c_{mn}({\bf r})$ obeys the Perelomov
 overcompleteness equation\cite{Per}
\begin{equation}
 \sum_{mn = - \infty}^{\infty} (-)^{m+n} c_{mn}({\bf r}) = 0
\label{eqOC}
\end{equation}
 Therefore functions $ c_{mn}({\bf r})$ are linearly dependent.

 The operators
 $T_{m{\bf a}} T_{n{\bf b}}$ in Eq. (\ref{eq0.1}) are considered as elements
 of the outer Kronecker product ${\rm G}_{\bf a} \otimes {\rm G}_{\bf b}$
 of the groups ${\rm G}_{\bf a}$ and ${\rm G}_{\bf b}$ of the
 magnetic translations along ${\bf a}$ and ${\bf b}$.
 It follows from Eq.~(\ref{eq0.2}) that 
 $T_{\bf a} T_{\bf b} = T_{\bf b} T_{\bf a}{\rm e}^{iab}
 = T_{\bf b} T_{\bf a}$. Therefore, $T_{m{\bf a}}$ and $T_{n{\bf b}}$ commute
 for arbitrary values of $m$ and $n$. By definition,\cite{Lax}
 the outer product consists of  operators
 $T_{m{\bf a}} T_{n{\bf b}}$  and does not include operators
 $T_{m{\bf a}+n{\bf b}}$ with $mn \neq 0$ which do not commute
 with the operators of the $T_{m{\bf a}}$ and $T_{n{\bf b}}$ types.
  Therefore, the Kronecker product
 ${\rm G}_{\bf a} \otimes {\rm G}_{\bf b}$
 forms an Abelian group of translations,
  and one can introduce a two-dimensional momentum $\bf k$
 and follow the Wannier procedure.\cite{Wann} 

 Transformation of the set $c_{mn}({\bf r})$ to the momentum representation results in Bloch functions:
 \begin{equation}
 \Psi_{\bf k}({\bf r}) = \sum_{mn= - \infty}^{\infty}
 c_{mn}({\bf r}) \exp(i{\bf k}{\bf R}_{mn})/
 \sqrt{\alpha \beta ~ \nu ({\bf k})}~.
\label{eq2}
\end{equation}
Functions $\Psi_{\bf k}({\bf r})$ with different values of $\bf k$ 
are orthogonal since they belong to different irreducible representations
 of the group ${\rm G}_{\bf a} \otimes {\rm G}_{\bf b}$.
 Values of $\bf k$ can be found from the boundary  conditions for
 $\Psi_{\bf k}({\bf r})$. The latter can be chosen either as 
 regular periodic boundary conditions
 with respect to the magnetic translations with 
 periods $L_{x}$ and $ L_{y}$~,
 or from ${\bbox \phi}$-periodic conditions when some twist
 ${\bbox {\phi}} = (\phi_{x},\phi_{y})$
 is added.\cite{NTW} Therefore, the components of  $\bf k$ can take values
\begin{equation}
 k_{x} = (2\pi s_{x} + \phi_{x})/L_{x}, ~~
 k_{y} = (2\pi s_{y} + \phi_{y})/L_{y}
\label{eq0}
 \end{equation}
 inside the Brillouin zone. Here  components
  $s_{x}$ and $ s_{y}$  of the vector ${\bf s} = (s_{x},~s_{y}) $
 are integers, $s_{x},~ s_{y} = 0,~ \pm 1, ...$~, taking $\alpha$ and $\beta$
 values, respectively. 
Normalization coefficient $\nu ({\bf k})$ is determined by the equation
\begin{equation}
 \nu ({\bf k}) = \sqrt{2\pi} \sum_{mn=-\infty}^{\infty}
  c_{mn}(0, 0)\cos({\bf k}{\bf R}_{mn})~.
\label{eq3}
\end{equation}
 The inverse Fourier transformation
\begin{equation}
 W_{mn}({\bf r}) = \sum_{\bf k}\Psi_{\bf k}({\bf r})~
 \exp(-i{\bf k}{\bf R}_{mn})/\sqrt{\alpha \beta}
\label{eq4}
\end{equation}
 results in a set of orthonormal  Wannier-type functions:
\begin{equation}
 W_{mn}({\bf r}) = \sum_{m'n'=-\infty}^{\infty}
 K_{ \bbox{\phi} }(m' - m,~ n' - n)~ c_{m'n'}({\bf r})~,
\label{eq5}
\end{equation}
 where
\begin{equation}
 K_{ \bbox{\phi} }(m, n) = {1\over {\alpha \beta}}
 \sum_{\bf k} \exp(i{\bf k}{\bf R}_{mn}) / \sqrt{\nu ({\bf k})}~.
\label{eq6}
\end{equation}
 The functions $ K_{ \bbox{\phi} }(m, n)$
 and $W_{mn}({\bf r})$ obey the equations
 $K_{\bbox{\phi}}(m + {\rm M}\alpha, n + {\rm N} \beta)
 = K_{\bbox{\phi}}(m, n) \exp\{i\phi({\rm M}, {\rm N})\}~,$ and
\begin{equation}
 W_{m+{\rm M}\alpha, n+{\rm N}\beta}({\bf r}) =
 W_{mn}({\bf r}) \exp\{ - i\phi({\rm M}, {\rm N}) \}~,
\label{eq0.3}
\end{equation} 
 where $\phi({\rm M}, {\rm N}) = {\rm M}\phi_{x}+{\rm N}\phi_{y}$~, and
 $\rm M$ and $\rm N$ are integers. Therefore, $|W_{mn}({\bf r})|$
 is periodic with periods $L_x$ and $L_y$ for any value of
 the flux ${\bbox \phi}$.

 The above equation for the Bloch functions $\Psi_{\bf k}({\bf r})$
 is known both in Landau\cite{BychR,HRtheta,Suth} and coherent
 function\cite{Fer90} basises, and the equation for $W_{mn}({\bf r})$
 was discussed more recently
 as applied to finite plaquettes.\cite{F95} These
 equations have their analogs in the
 mixed $kq$ representation.\cite{Zak}

  Functions $\Psi_{\bf k}({\bf r})$ are orthogonal and, therefore, 
 linearly independent. If  they exist for any momenta
 ${\bf k} = (k_{x}, {k_y})$, defined by Eq.~(\ref{eq0}), then the set $\Psi_{\bf k}({\bf r})$
 is complete. Hence,  the set $W_{mn}({\bf r})$ is complete, too.
 However, the properties of one of the functions $\Psi_{\bf k}({\bf r})$,
 namely the function $\Psi_{{\bf k}_{0}}({\bf r})$, where ${\bf k}_{0}$
is the corner of the Brillouin zone, need a special deliberation.
 At this point
 $k_{x}a = \pm ~k_{y}b = \pm ~\pi$, and the exponent in Eq.~(\ref{eq2})
 equals $(-)^{m+n}$. Therefore, the numerator of 
 $\Psi_{{\bf k}_{0}}({\bf r})$ turns into zero because of Eq.~(\ref{eqOC}).
 If one puts $\Psi_{{\bf k}_{0}}({\bf r}) = 0$ and excludes this function from
 the set $\Psi_{\bf k}({\bf r})$, the set becomes incomplete since it
 lacks a function of the translation symmetry of ${\bf k}_{0}$. This
 problem was emphasized by Thouless.\cite{Thou}

 Since functions $\Psi_{\bf k}({\bf r})$ are normalized, the denominator
 of $\Psi_{{\bf k}_{0}}({\bf r})$ also turns into zero, $\nu({\bf k}_{0})=0$.
 Indeed, for ${\bf k}={\bf k}_{0}$ the cosine in Eq.~(\ref{eq3}) equals 
 $\cos({\bf k}_{0}{\bf R}_{mn}) = (-)^{m+n}$, and therefore
 $\nu({\bf k}_{0})=0$ due to Eq.~(\ref{eqOC}). Function $\nu({\bf k})$ is
 shown in Fig.~{\ref{fig1}}
 for a square lattice, $a = b = \sqrt{2\pi}$. One can
 see that $\nu({\bf k})$ is positive,
 $\nu({\bf k}) > 0$, inside the Brillouin zone and reaches its minima,
 $\nu({\bf k}_{0})=0$, in the corners of it. Both the numerator
 and denominator of $\Psi_{\bf k}({\bf r})$ have the order of magnitude of
  $|{\bbox \zeta}|$
 when ${\bbox \zeta}\equiv {\bf k} - {\bf k}_{0} \rightarrow 0$.
 The leading term in the numerator of $\Psi_{\bf k}({\bf r})$
 depends on $\varphi_{\bbox \zeta}$~,
 the azimuth of ${\bbox \zeta}$~, as it is shown below; cf. Eq.~(\ref{eq11}).
 On the contrary, the function  $\sqrt{\nu({\bf k})}$, which is the
 denominator  of $\Psi_{\bf k}({\bf r})$, 
 is isotropic near ${\bf k}_0$ for a square lattice.
 Therefore, the function $\Psi_{\bf k}({\bf r})$ retains the dependence
 on $\varphi_{\bbox \zeta}$ even in the limit ${\bbox \zeta}\rightarrow 0$.
 Hence, it is singular at the point ${\bf k}={\bf k}_0$,
 and the limit $\Psi_{{\bf k}\rightarrow {{\bf k}_{0}}}$ does not exist.
 
 To find the form of $\Psi_{\bf k}({\bf r})$ in the limit
 ${\bf k} \rightarrow {\bf k}_{0}$~, one can expand the numerator of
 $\Psi_{{\bf k}_{0}+{\bbox \zeta}}({\bf r})$ in ${\bbox \zeta}$ and
 take into account Eq.~(\ref{eqOC}).  Then
\begin{equation}
 \Psi_{{\bf k}_{0}+{\bbox \zeta}}({\bf r}) =
 {i\over { [\alpha \beta \nu({\bf k})]^{1/2} }}
 \sum_{mn= - \infty}^{\infty} (-)^{m+n}~
 {\bbox \zeta}\cdot {\bf R}_{mn}~c_{mn}({\bf r})~.
\label{eq7}
\end{equation}
 If one introduces complex variables $z=x+iy$ and
 $Z_{mn} = X_{m} + i Y_{n}$, the condition (\ref{eqOC}) takes the form:
\begin{equation}
\sum_{mn= - \infty}^{\infty} (-)^{mn+m+n}
 \exp\{ - {1\over 4}{\bf R}_{mn}^{2} + {1\over 2}z{\bar Z}_{mn} \} = 0~.
\label{eq8}
\end{equation}
 Here and below complex conjugate variables are designated by bars.
 Since this equation is valid for arbitrary values of $z$ and the sum 
 converges exponentially, one can take the derivative  over $z$:
\begin{equation}
\sum_{mn= - \infty}^{\infty} (-)^{m+n}~{\bar Z}_{mn}~c_{mn}({\bf r}) = 0~.
\label{eq9}
\end{equation}
 If one takes advantage of the relation ${\bbox \zeta}\cdot{\bf R}_{mn}=
 ({Z}_{mn}{\bar \zeta} + {\bar Z}_{mn}{\zeta})/2$,
 where $\zeta = {\zeta}_{x}+i{\zeta}_{y}$~, and plugs Eq.~(\ref{eq9})
 into Eq.~(\ref{eq7}), the latter takes the form:
\begin{equation}
 \Psi_{{\bf k}_{0}+{\bbox \zeta}}({\bf r}) \approx
 { {i/2} \over { [\alpha \beta \nu({\bf k})]^{1/2} }}~ {\bar \zeta}
 \sum_{mn= - \infty}^{\infty} (-)^{m+n}
 Z_{mn}~c_{mn}({\bf r})~.
\label{eq10}
\end{equation}
Since ${\nu({\bf k})}^{1/2} \propto |\zeta |$ for $|\zeta |\rightarrow 0$,
 Eq.~(\ref{eq10}) can be rewritten as
\begin{equation}
  \Psi_{{\bf k}_{0}+{\bbox \zeta}}({\bf r}) \approx
 {\rm e}^{-i{\varphi}_{\bbox \zeta} }~\Phi_{{\bf k}_{0}}({\bf r})~,
\label{eq11}
\end{equation}
 where $\Phi_{{\bf k}_{0}}({\bf r})$ is a regular function of $\bf r$
 possessing the symmetry of the point ${\bf k}_0$.

 Therefore, nonanalitic function $ \Psi_{{\bf k}_{0}+{\bbox \zeta}}({\bf r})$
 factors near the singular point into 
 the  product of two functions, ${\rm e}^{-i{\varphi}_{\bbox \zeta} }$
 and $\Phi_{{\bf k}_{0}}({\bf r})$~. The first factor absorbs the
 nonanalytic dependence of 
 $ \Psi_{{\bf k}_{0}+{\bbox \zeta}}({\bf r})$
 on ${\bbox \zeta}$, whereas the second factor does not depend
 on ${\bbox \zeta}$ and possesses the translational symmetry of the
 point ${\bf k}_0$. Function $\Phi_{{\bf k}_{0}}({\bf r})$, taken
 with an arbitrary phase factor, can be used as a Bloch function with
 the ${\bf k}_0$ symmetry. Therefore, Bloch functions are defined
 for all $\bf k$ values. This statement concludes the proof of the
 completeness of the set $ \Psi_{\bf k}({\bf r})$. The set 
 $W_{mn}({\bf r})$ obtained from it by the orthogonal transformation
 of Eq.~(\ref{eq4}) is also complete.

 There exists another way to construct the function $\Phi_{{\bf k}_{0}}({\bf r})$.
 It is based on the properties of Bloch functions constructed from different
sets of localized  orbitals. Instead of
 $c_{00}(r)$, one can use the function $c_{00}^{(1)}({\bf r})=
 zc_{00}(r)/\sqrt{2}$
 to generate the sets of orbitals $c_{mn}^{(1)}({\bf r})$
 and Bloch functions $\Psi_{\bf k}^{(1)}({\bf r})$. The sets
  $c_{mn}({\bf r})$ and $c_{mn}^{(1)}({\bf r})$ belong to the LLL.
 Both sets are overcomplete  and can be expanded one in another.
 Consequently, functions $\Psi_{\bf k}({\bf r})$ and
 $\Psi_{\bf k}^{(1)}({\bf r})$ can differ only in $\bf r$
 independent phase factors for arbitrary value of $\bf k$.
  Boon {\it et al.} have shown\cite{Boon}
 that ${\bf k}_0$ is a regular point for the set 
  $\Psi_{\bf k}^{(1)}({\bf r})$.  
 Therefore, one can use
 $\Psi_{{\bf k}_{0}}^{(1)}({\bf r})$
 as a function $\Phi_{{\bf k}_{0}}({\bf r})$.
 The shape of the function $W_{mn}({\bf r})$ depends on
 the choice of the phase of $\Phi_{{\bf k}_{0}}({\bf r})$.

 The properties of Bloch functions discussed in the previous paragraph
  can be also understood from a more general point of view.  It follows from
 Eqs.~(\ref{eq1}) and (\ref{eq8}) that functions $c_{mn}^{(1)}({\bf r})$
 can be obtained as
\[
 c_{mn}^{(1)}({\bf r}) = \sqrt{2}~ \left(\partial/{\partial {\bar Z}_{mn}}
 - {1\over 4}Z_{mn}\right)~ c_{mn}({\bf r})~.
\]
Therefore, functions $c_{mn}^{(1)}({\bf r})$ and $c_{mn}({\bf r})$ belong to 
the same Landau level because of  the existence of continuous
 group of magnetic translations. The same is true for the functions
 $c_{mn}^{(q)}({\bf r})$ originating from the function
 $c_{00}^{(q)}({\bf r}) \propto z^{q}c_{00}({\bf r})$, where $q$ is
 an arbitrary integer. Bloch functions $\Psi_{\bf k}^{(q)}({\bf r})$,
 constructed from them, are completely determined by the translational
 symmetry up to the phase factors depending only on $q$ and $\bf k$.
 Therefore, all functions $\Psi_{\bf k}^{(q)}({\bf r})$ with the same
 value of the momentum $\bf k$ and different values of $q$ 
 coincide up to these phase factors. Singularities of functions
 $\Psi_{\bf k}^{(q)}({\bf r})$ ensure the existence of the Hall 
 current,\cite{Thou} hence, they are present for each set of 
 functions $\Psi_{\bf k}^{(q)}({\bf r})$ and can not be eliminated.
 However, the number of singular points and their positions in the
 Brillouin zone change depending on $q$.\cite{Boon}

 It is known that the overcompleteness equation (\ref{eqOC}) is
 related to the properties of Jacobi $\vartheta(u|\tau)$ functions.\cite{Per,Thou}
 Eq.~(\ref{eq3}) for $\nu({\bf k})$ takes a simple form when rewritten in terms
 of these functions. If one performs the summation over $n$ in Eq.~(\ref{eq3})
 using the Jacobi imaginary transformation of
 theta functions,\cite{Erd} $\nu({\bf k})$ acquires the form:
\begin{equation}
 \nu({\bf k}) = {a\over {\sqrt{\pi}}}~ {\rm e}^{-k_{y}^{2}}~
 {\vartheta}_{3}\left({a\over 2\pi}k_{+} \mid i{a\over b}\right)~
 {\vartheta}_{3}\left({a\over 2\pi}k_{-} \mid i{a\over b}\right) ~,
\label{eq12}
\end{equation}
where $k_{\pm} = k_{x}\pm ik_{y}$. The zeros of $\vartheta_{3}(u|\tau)$
 can be found from the condition $u=({\rm M}+{1\over 2}) +
 ({\rm N}+{1\over 2}) \tau$, where ${\rm M}$ and ${\rm N}$ are
 integers, and one immediately recovers that the corner of the Brillouin
 zone,
 ${\bf k}_{0} = (\pm {\pi /a},~ \pm {\pi /b})$~,
 is a  zero of $\nu({\bf k})$.

 \section{Wannier functions in the infinite plane}
\label{sec:plane}

 Equations (\ref{eq5}) and (\ref{eq6}) can be used to find Wannier
 functions $w_{mn}({\bf r})$ localized in the infinite $\bf r$ plane.
 In the limit $\alpha, \beta \rightarrow \infty$, the sum in Eq.~(\ref{eq6})
transforms into the integral over the Brillouin zone:
\begin{equation}
 K_{\infty}(m, n) = ab ~\int_{(B.Z.)} ~{d{\bf k}\over {(2\pi)^{2}}}~
 {\exp(i{\bf k}{\bf R}_{mn})\over\sqrt{ {\nu ({\bf k})}}}~.
\label{eq3.1}
\end{equation}
 The kernel $ K_{\infty}(m, n)$ is obviously independent of $\bbox \phi$.
 
 The asymptotic behavior of $K_{\infty}(m, n)$ for $|m|, |n|\gg 1$
 can be found analytically. It is determined by the behavior of the
 integrand near its pole, i.e., by the expansion of $\nu ({\bf k})$
 near ${\bf k}_{0}$. For a square lattice, $a=b=\sqrt{2\pi}$,
 this expansion has the form
\begin{equation}
 \nu({\bf k}_{0}+{\bbox \zeta}) \approx \gamma a^{2} {\bbox \zeta}^{2}/2~,
\label{eq3.2}
\end{equation}
where $\gamma \approx 0.5814$ is given by the series:
\begin{equation}
\gamma = - \sum_{mn=-\infty}^{\infty} (-)^{mn+m+n} m^{2}
 \exp [- {\pi \over 2}(m^{2} + n^{2})].
\label{eq3.3}
\end{equation}
 Substituting (\ref{eq3.2})  into (\ref{eq3.1}) results in the leading term
 of the expansion of $K_{\infty}(m, n)$ in $R_{mn}^{-1}$:
\begin{equation}
 K_{\infty}(m, n) 
 \approx (-)^{m+n}~ 2~\sqrt{\pi \over \gamma}~ 
 \int_{-\infty}^{\infty} {d{\bbox \zeta}\over {(2\pi)^{2}} }
 ~{ {\exp(i{\bbox \zeta}{\bf R}_{mn}) }\over {\zeta} } =
 { (-)^{m+n} \over {\sqrt{\pi \gamma}~ {R}_{mn}} }~.
\label{eq3.4}
\end{equation}
  The next term of this expansion  falls off as
 $R_{mn}^{-3}$. An equation equivalent to 
 $K_{\infty}(m, n) \propto (-)^{m+n}/R_{mn}$ was derived by Sen and
 Chitra\cite{Sen} using a different procedure.

 The asymptotic behavior of $w_{mn}({\bf r})$ for large $|{\bf r}-{\bf R}_{mn}|$
 values follows from Eqs.~(\ref{eq5}) and (\ref{eq3.4}).
 The right hand side of  the equation
\begin{equation}
 w_{00}({\bf r}) = \sum_{mn=-\infty}^{\infty}
  K_{\infty}(m, n)~c_{mn}({\bf r})
\label{eq3.5}
\end{equation} 
 for the function $w_{00}({\bf r})$  includes the
 product of the kernel $K_{\infty}(m, n)$, whose denominator depends
 on $m$ and $n$ slowly, and the factor $c_{mn}({\bf r})$, which depends
 on $m$ and $n$ exponentially for a fixed value of $\bf r$. One can use
 the asymptotic form of the kernel $K_{\infty}(m, n)$, Eq.~(\ref{eq3.4}),
 and neglect the higher order corrections to it. If one substitutes  the expansion
 \[
 R_{mn}^{-1} \approx r^{-1} [1 - {\bf r}\cdot ({\bf R}_{mn}-{\bf r})/r^{2} ]
\]
 into Eq.~(\ref{eq3.5}), the first vanishes because
 of the Perelomov identity, Eq.~(\ref{eqOC}), and the second term results in 
\begin{equation}
 w_{00}({\bf r}) \approx
 -~ {{\bf r} \over {\sqrt{\pi \gamma}~r^{3}}}~
 \sum_{mn=-\infty}^{\infty}
 (-)^{m+n}~({\bf R}_{mn} - {\bf r})~c_{mn}({\bf r})~.
\label{eq3.6}
\end{equation} 
 The first factor in Eq.~(\ref{eq3.6}) falls off
 as $r^{-2}$, whereas the second  is a periodic
 function of $\bf r$. If one plugs ${\bf r}={\bf R}_{m^{'}n^{'}}$ in the
 second factor and applies Eq.~(\ref{eq1}), this factor takes the form:
 \[
 (-)^{m^{'}+n^{'}+m^{'}n^{'}}  \sum_{mn=-\infty}^{\infty}
 (-)^{m+n+mn} ~{\bf R}_{mn}~ c_{00}({\bf r})~.
 \]
 The sum is equal to zero because of symmetry arguments, therefore, all lattice
 sites, ${\bf r}={\bf R}_{mn}$, are zeros of the asymptotic expansion 
 (\ref{eq3.6}) of the   function $w_{mn}({\bf r})$.

 It was shown by Kohn\cite{Kohn} that the rate of the falloff of an exponentially
 decaying Wannier function is determined by the distance of the
 singular point in the complex momentum plane from the real axis.
  In the problem of magnetic Wannier functions unavoidable singularities
  exist in the real $(k_{x}, k_{y})$ plane. These singularities result in the
 power-law falloff of the Wannier functions.

 The asymptotic
 expansion of $K_{\infty}(m, n)$, Eq.~(\ref{eq3.4}), is accurate
 up to the values  $|m|, |n| \approx 1$, e.g., the deviation of
 $K_{\infty}(1, 0) \approx - 0.288$ from its approximate value following from
 Eq.~(\ref{eq3.4})  is only about 2\%. The coefficient
 $K_{\infty}(0, 0) \approx 1.241$ is much larger than all the coefficients
 $K_{\infty}(m, n)$  with $m, n \neq 0$. One can subtract this large term and
 rewrite Eq.~(\ref{eq3.5}) in the form
\begin{equation}
w_{00}({\bf r}) - c_{00}({\bf r}) =
 \sum_{mn=-\infty}^{\infty} ~(-)^{m+n}~{\Delta}(m^{2}, n^{2})~c_{mn}({\bf r})~,
\label{eq3.7}
\end{equation} 
 The kernel ${\Delta}~(m^{2}, n^{2})$ is numerically small and can
 be considered as a smooth function of $m$ and $n$. This allows
 the application of  the arguments employed when deriving Eq.~(\ref{eq3.6}).
 Then the leading term in the right hand side of Eq.~(\ref{eq3.7})
 vanishes because of the overcompleteness condition of Eq.~(\ref{eqOC}).
 Therefore $w_{00}({\bf r}) \approx  c_{00}({\bf r})$, which means
 that the difference between the functions $w_{00}({\bf r})$ and
 $c_{00}({\bf r})$ is expected to be small in the region where these
 functions are large. In the large $r$ region, where these functions
 are small, $w_{00}({\bf r})$ dominates and Eq.~(\ref{eq3.6}) should be used.

Numerical results support all of the above conclusions. 
 The infinite-plane function $w_{00}({\bf r})$ is shown in Fig.~\ref{fig2}.
 Fig.~\ref{fig2}a provides a detailed comparison of the shapes of the function
 $|w_{00}({\bf r})|^2$ and the Gaussian function $c_{00}(r)^2$ in
 the region of $|x|,~|y| \leq 1.5 \times \sqrt{2\pi}$.
 These  functions
 are plotted along the $x$ axis and along the diagonal, $x = y$, on the
 right and left hand sides of the figure, respectively.
 The function $w_{00}({\bf r})$ is real on these lines.
 One can see that
  $w_{00}({\bf r})^2$ and  $c_{00}(r)^2$ are very close to each other
 in both directions. The function 
 $w_{00}({\bf r})^2$ shows small
 anisotropy:
it is slightly elongated in the $x = y$ direction and squeezed
 in the $x$ direction compared with $c_{00}^{2}(r)$.
  It is a remarkable property of the function
 $|w_{00}({\bf r})|^2$ that it is very small in the points
${\bf r} = \sqrt{2\pi}(1,~0)$ and $ \sqrt{2\pi}(1,~1)$.
 These points are the lattice sites adjacent to the origin.
 The suppression of $|w_{00}({\bf r})|^2$  at the lattice sites
 of the two first coordination spheres of the von Neumann lattice 
 results in the
 squeezing of the central peak and
 a strong localization of $|w_{00}({\bf r})|^2$.  Consequently,
 the central region contributes about 95\% to the normalization
 integral. The behavior of $w_{00}({\bf r})$ along the $x$ axis is shown in
 Fig.~\ref{fig2}b
 over a wide region of $x$ values. 
 It is seen that $w_{00}(x, 0)$ oscillates
 and decreases with $x$. For $x\rightarrow \infty$,
 the oscillation
 amplitudes decrease as $x^{-2}$ and zeros of $w_{00}(x, 0)$
 approach multiples of the lattice period
 $\sqrt{2\pi}$ in agreement with  Eq.~(\ref{eq3.6}).
 The exact $w_{00}({\bf r})$ curve is almost indistinguishable from
 its asymptotic shape in the entire oscillatory region.

 Therefore, functions $w_{mn}({\bf r})$ show a two-scale behavior.
 The function $w_{00}({\bf r})$ is large and shows only minor deviations
 from the Gaussian shape inside the central cell, 
 but it is small, oscillates, and falls off according the
 $r^{-2}$ law in the asymptotic region.

  The set $w_{mn}({\bf r})$ is orthogonal and complete by arguments
 of Sec.~\ref{sec:equations}. Therefore, any function which is invariant 
 under the unitary transformations of the basis can be calculated in the
 Wannier representation. For instance, a straightforward calculation
  based on the representation of the functions $w_{mn}({\bf r})$
 through the kernel $K_{\infty}(m, n)$ and coherent functions
 $c_{mn}({\bf r})$ shows that the function
\begin{equation}
 C_{\infty}({\bf r}, {\bf r'})
 = \sum_{mn=-\infty}^{\infty}{\bar w}_{mn}({\bf r}) w_{mn}({\bf r'})
 \label{eqa}
\end{equation}
 is equal to the  expression
\begin{equation}
 C_{\infty}({\bf r}-{\bf r'})=(1/2\pi)\exp\{-({\bf r}-{\bf r'})^2/4\}
 \exp\{{i\over 2}~\hat{{\bf z}}\cdot ({\bf r}\times {\bf r'}) \}.
 \label{eqb}
\end{equation}
 which can be also derived in the Landau representation. The last factor 
in Eq.~(\ref{eqb}) is gauge dependent.
 Therefore, the continuous symmetry of the magnetic translation group
 is recovered due to the completeness of the system $w_{mn}({\bf r})$.

\section{Some results for  toroidal geometry}
\label{sec:toroidal}

 Only few results can be obtained for
 Wannier functions $W_{mn}({\bf r})$, Eq.~(\ref{eq4}),
 analytically. However, numerical
 methods allow one to study the dependence of $W_{mn}({\bf r})$
 on the plaquette size $(\alpha, \beta)$ and the twist ${\bbox \phi}$,
 and to check the self-consistency of the procedure. A number of results
 was obtained by Ferrari.\cite{F95,Fer90} The singular behavior of
 $\Psi_{\bf k}({\bf r})$ near ${\bf k}={\bf k}_0$ was studied
 in Sec.~\ref{sec:equations}.

 In Fig.~\ref{fig3} the square of the modulus of the function
 $W_{00}({\bf r})$ as well as the real and imaginary parts 
 of $W_{00}({\bf r})$ are shown
 for a square lattice with $\alpha=\beta=3$ and ${\bbox {\phi}} =0$.
 It is seen that $|W_{00}({\bf r})|^2$ is nearly isotropic
 and well localized in the area of about $\sqrt{2\pi}$  near the origin.
 The shape of this function is  close to the data\cite{F95}
 for a triangular lattice. Re\{$W_{00}({\bf r})\}$  is also rather isotropic
 and well localized, whereas Im\{$W_{00}({\bf r})\}$ is small and highly 
 anisotropic in the same region. The shape of the function
 Im\{$W_{00}({\bf r})\}$ can be understood if one takes into account
 that, because of Eqs.~(\ref{eq1}) and (\ref{eq5}), the expansion of
 Im\{$W_{00}({\bf r})\}$ in the powers of $z$ starts with the term
 Im\{$W_{00}({\bf r})\} \propto {\rm Im}\{z^{4}\}$. All
 lower-power terms cancel for a square lattice.  

 It is seen in Fig.~\ref{fig3} that all three functions possess full
 symmetry of the square lattice. For an odd-odd plaquette this high-symmetry
 shape appears only for the twist ${\bbox \phi} = 0$. When twist 
 increases, the shape of $|W_{00}({\bf r})|^2$ changes and becomes asymmetric.
 The changes are moderate near the maximum of the surface 
 $|W_{00}({\bf r})|^2$ but are considerably larger in its lower part.
 Dependence of the  function $W_{00}({\bf r})$  on ${\bbox \phi}$ becomes discontinuous when 
 one of the points determined
 by Eq.~(\ref{eq0}) passes through the corner of the
 Brillouin zone ${\bf k}_0$, 
  since in the vicinity of ${\bf k}_0$
 the function $\Psi_{\bf k}({\bf r})$  critically depends on the phase
 $\varphi_{\bbox \zeta}$ of ${\bbox \zeta}={\bf k}-{\bf k}_0$ as it follows
 from Eq.~(\ref{eq11}).

 More generally,
 the function $|W_{mn}({\bf r})|^2$ depends on $\bbox \phi$ discontiniously when
 i)  both $\alpha$ and $\beta$ are odd and $\phi_{x}=\phi_{y}=\pi$,
 ii) both $\alpha$ and $\beta$ are even and ${\bbox \phi}=0$,
 and iii) $\alpha$ is odd, $\beta$ is even, and $\phi_{x}=\pi$ and
 $\phi_{y}=0$.

 Toroidal Wannier functions constructed by magnetic translations 
 of infinite-plane functions $w_{mn}({\bf r})$
 have more stable shape than $W_{mn}({\bf r})$. This approach will be discussed 
 in more detail elsewhere.

 It was shown in Sec.~\ref{sec:equations} that despite of 
 the singular dependence  of $\Psi_{\bf k}({\bf r})$ and
  $W_{mn}({\bf r})$ on the twist ${\bbox \phi}$ in the vicinity
 of ${\bf k}_0$, these sets of functions remain complete at any
 values of ${\bbox \phi}$. The completeness can be checked by comparing
 the results obtained in  the  $W_{mn}({\bf r})$ lattice  basis
 with those in Landau function\cite{YHL} basis.
 For example, the function
\begin{equation}
 C({\bf r}, {\bf r'})
 = \sum_{mn} {\bar W}_{mn}({\bf r}) W_{mn}({\bf r'})~,
 \label{eq4.1}
\end{equation}
 which is similar to the function $C_{\infty}({\bf r}, {\bf r}^{'})$,
 Eq.~(\ref{eqa}), 
 does not depend on a specific choice of a
 complete basis set of the LLL.
 We have calculated the right hand side of Eq.~(\ref{eq4.1})
 for finite plaquettes  using both
 $W_{mn}({\bf r})$  and Landau functions as given in Ref. 2.
 Gauge independent parts of $C({\bf r}, {\bf r'})$ in both representations
 coincide with the accuracy
of our computations. They were performed
for the values of $\alpha$ and  $\beta$ up to 5.
 For the finite plaquettes  the electron density
 $\rho({\bf r}) = C({\bf r},{\bf r})$
 oscillates with ${\bf r}$ near its continuum
 limit $\rho_{\infty} = 1/{2\pi}$. The amplitude of  oscillations decreases
 rapidly with increasing  $\alpha$ and $\beta$. The number of oscillations
 in $x$ and $y$ directions equals  the number of fluxes
 $\alpha \beta$ per a plaquette rather than
 the individual values of $\alpha$ and $\beta$, which are the number of
 lattice sites in corresponding directions. This property was found
 analytically in the Landau representation by Sutherland.\cite{Suth} It
 indicates  disappearance
 of the pattern of the flux lattice and restoration of the 
 symmetry of the underlying problem as it is expected since both periodic
 functions $W_{mn}({\bf r})$ and Landau functions form  complete sets.
 The completeness of the $W_{mn}({\bf r})$ set is in agreement with
 the statements by Ferrari.\cite{F95}
 
\section{Conclusion}
\label{sec:conclusion}

 The construction of infinite-plane localized magnetic Wannier-type functions
 was considered  a challenging problem for a long time. It was shown that
 these functions, if they do exist, are subject to rigid restrictions.
 Three properties of a set of localized magnetic orbitals are 
 incompatible: completeness, orthogonality, and exponential falloff.\cite{Zak}
 Moreover, for systems supporting a Hall current magnetic orbitals
 should fall off no faster than  $r^{-2}$ .\cite{Thou}
 The set of  Gaussian coherent states on a
 von Neumann lattice violates these criteria because of the
 overcompleteness relation,\cite{Per} which is a single linear constraint
 relating an infinite set of orbitals. We believe that the complete set of
 orthogonal two-scale orbitals $w_{mn}({\bf r})$ studied in this paper
 is the best compromise between the requirement of optimal localization
 and the inevitable restrictions on the degree of localization. Function
 $w_{00}({\bf r})$ centered at the origin i) is large and possesses nearly
 Gaussian shape inside the region of the size of about $\sqrt{2\pi}l$
 making a dominant contribution to the normalization integral, and
 ii) is small and falls off with a critical exponent $r^{-2}$ outside
 this region. It is a striking property of the theory that the
 overcompleteness condition, Eq.~(\ref{eqOC}),
 emerges and plays a crucial role at all
 stages of the derivation and study of the Wannier orbitals
 $w_{mn}({\bf r})$. 

 We expect that $w_{mn}({\bf r})$ orbitals form a convenient basis
 both for analytical and numerical calculations.

\acknowledgments

 We are grateful to  M. I. D'yakonov,  B. I. Korenblum,
  E. V. Tsiper, and Y.-S. Wu for suggestive
discussions, to Bill Sutherland for providing us with his unpublished data,
 to J. Zak for  useful correspondence, and to R. Ferrari and D. Sen
 for sending us  references to their papers after our preprint
 appeared in the e-print  archive. 
  We also acknowledge the support of the
  QUEST of the UCSB, subagreement KK3017,
 and of the SDSC where part of the calculations were performed.

\begin{figure}
 \caption{Normalization factor $\nu({\bf k})$ for a square lattice.}
\label{fig1} 
\end{figure}

\begin{figure}
 \caption{(a) Dependence of the functions  $|w_{00}({\bf r})|^2$
 and $c_{00}(r)^2$ on $r$
 in the directions of the principal axes $x$ and \protect$y$
 (right) and the diagonals $x = \protect\pm ~y$ (left). Solid line -
 $|w_{00}({\bf r})|^2$,
 dashed line - Gaussian function $c_{00}(r)^2$.
(b) Oscillatory dependence of the
 function $w_{00}(x, 0)$ on $x$.
 The exact solution , asymptotic solution, and envelope
 function proportional to $ x^{-2}$ are shown in
 the region $x\protect\geq \protect\sqrt{2\pi} l$
by solid, dotted and dashed lines, respectively.}
\label{fig2}
\end{figure}

\begin{figure}
 \caption{  Shape of the function $W_{00}({\bf r})$ for a square
 lattice with $\alpha=\beta=3$ and
${\protect\bbox {\phi}} =0$.
 (a) - $|W_{00}({\bf r})|^2$, (b) - Re\protect\{$W_{00}({\bf r})$\protect\},
 (c) - Im\protect\{$W_{00}({\bf r})$\protect\}.}
\label{fig3}
\end{figure}

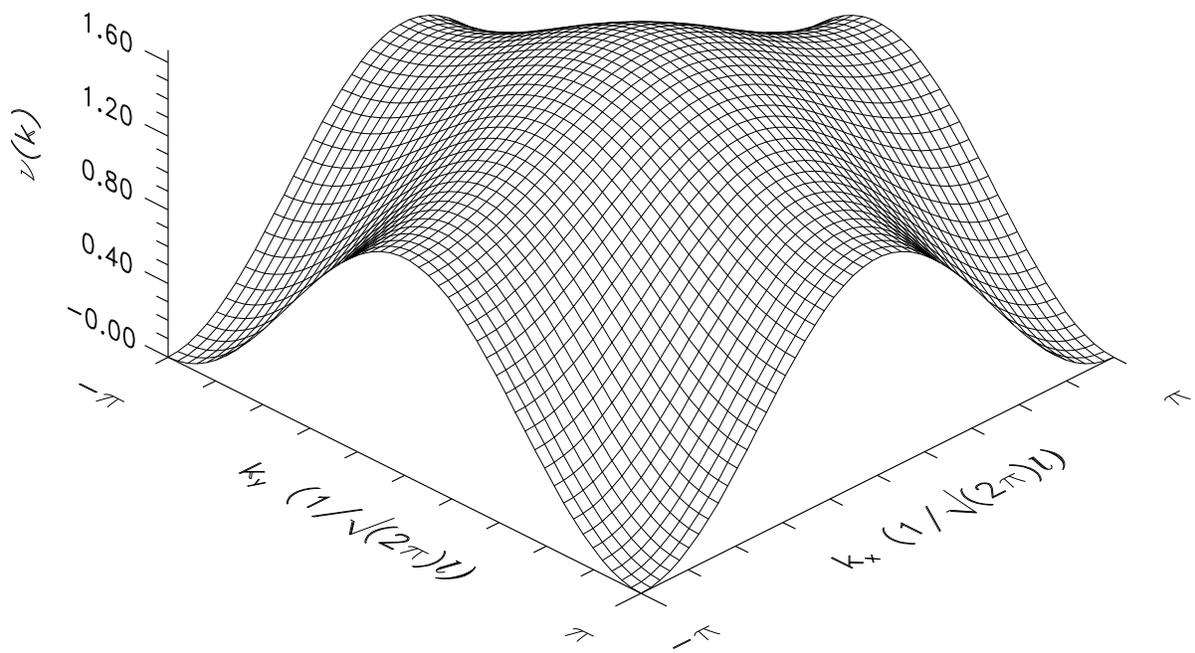
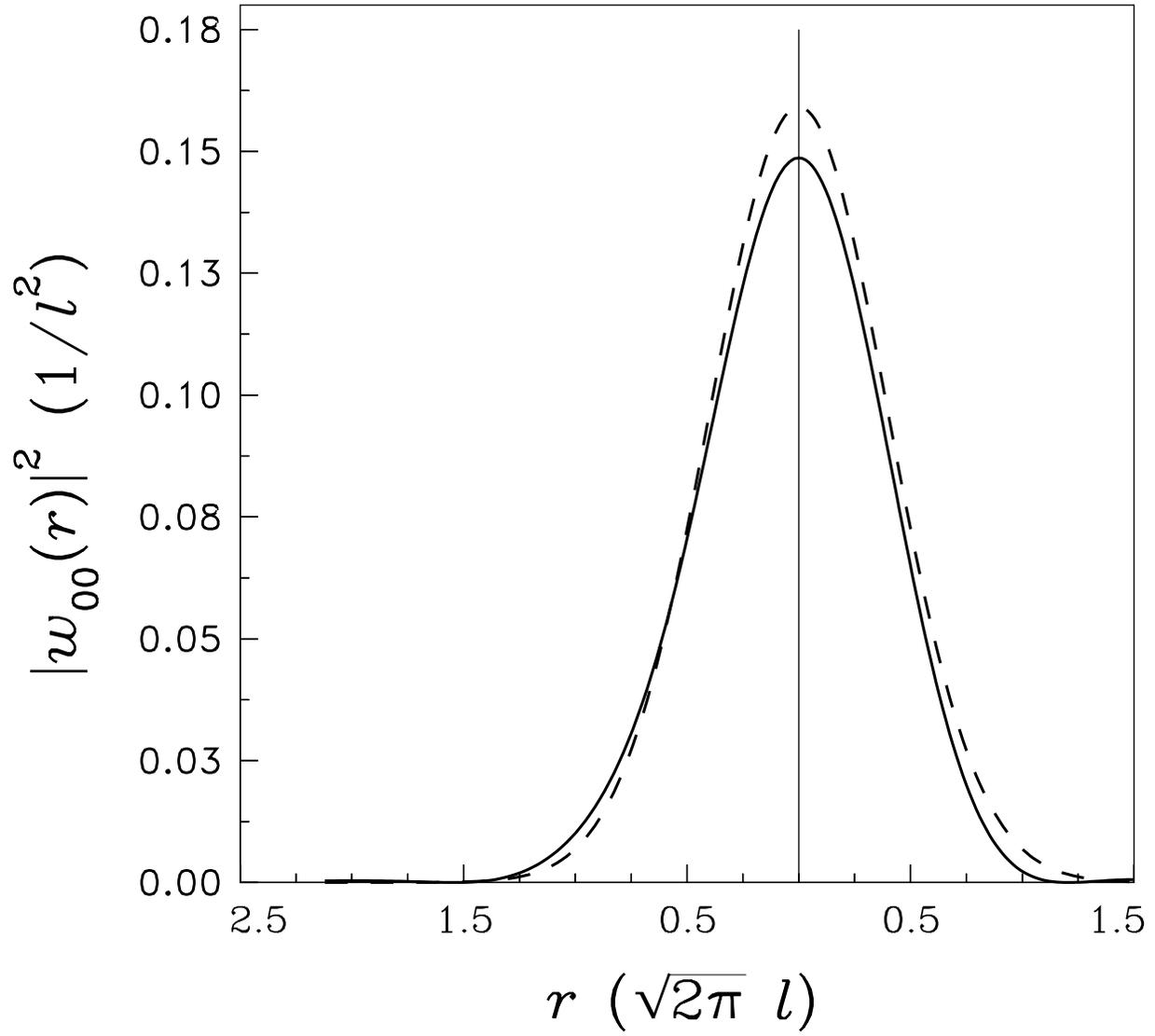
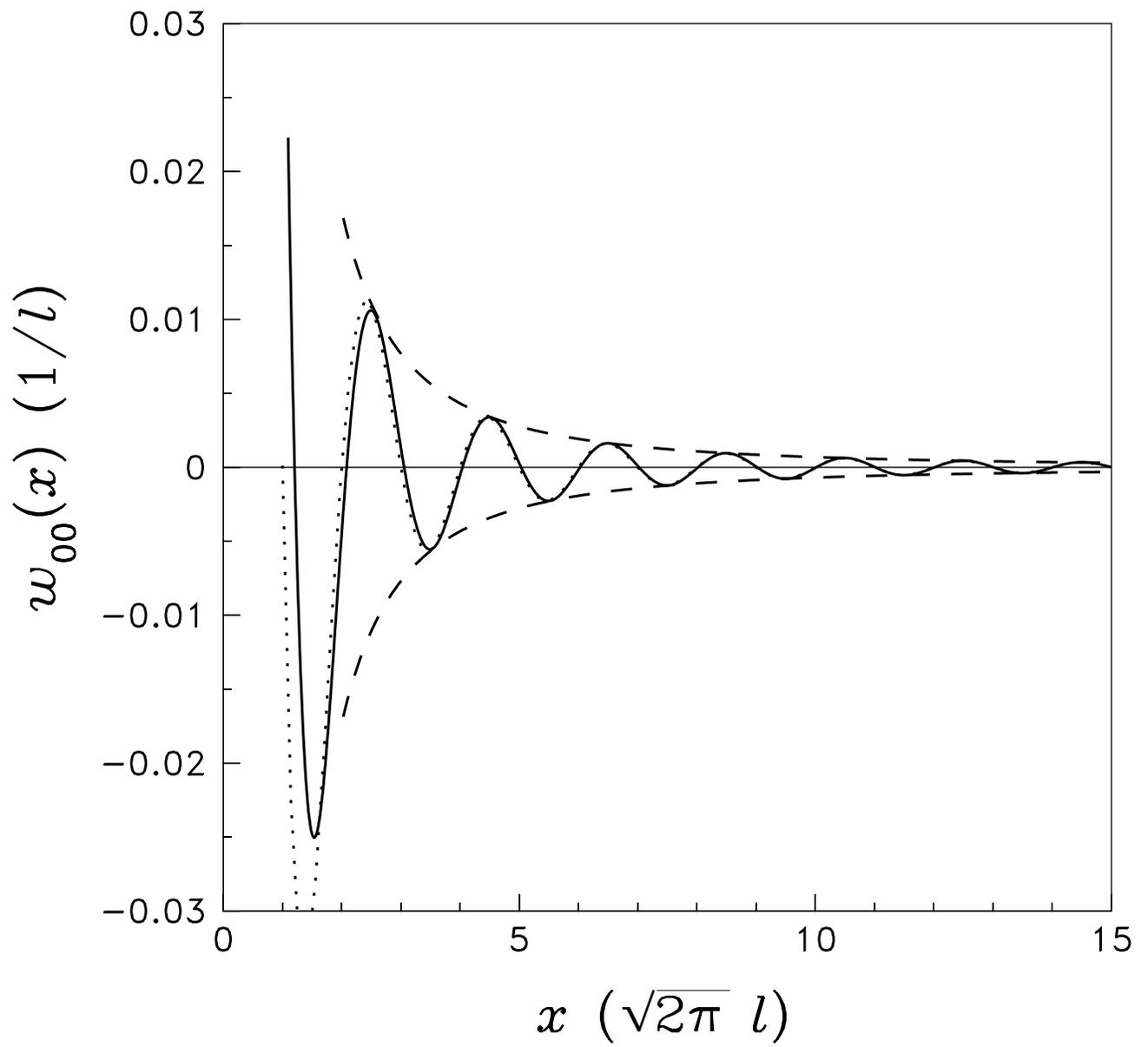
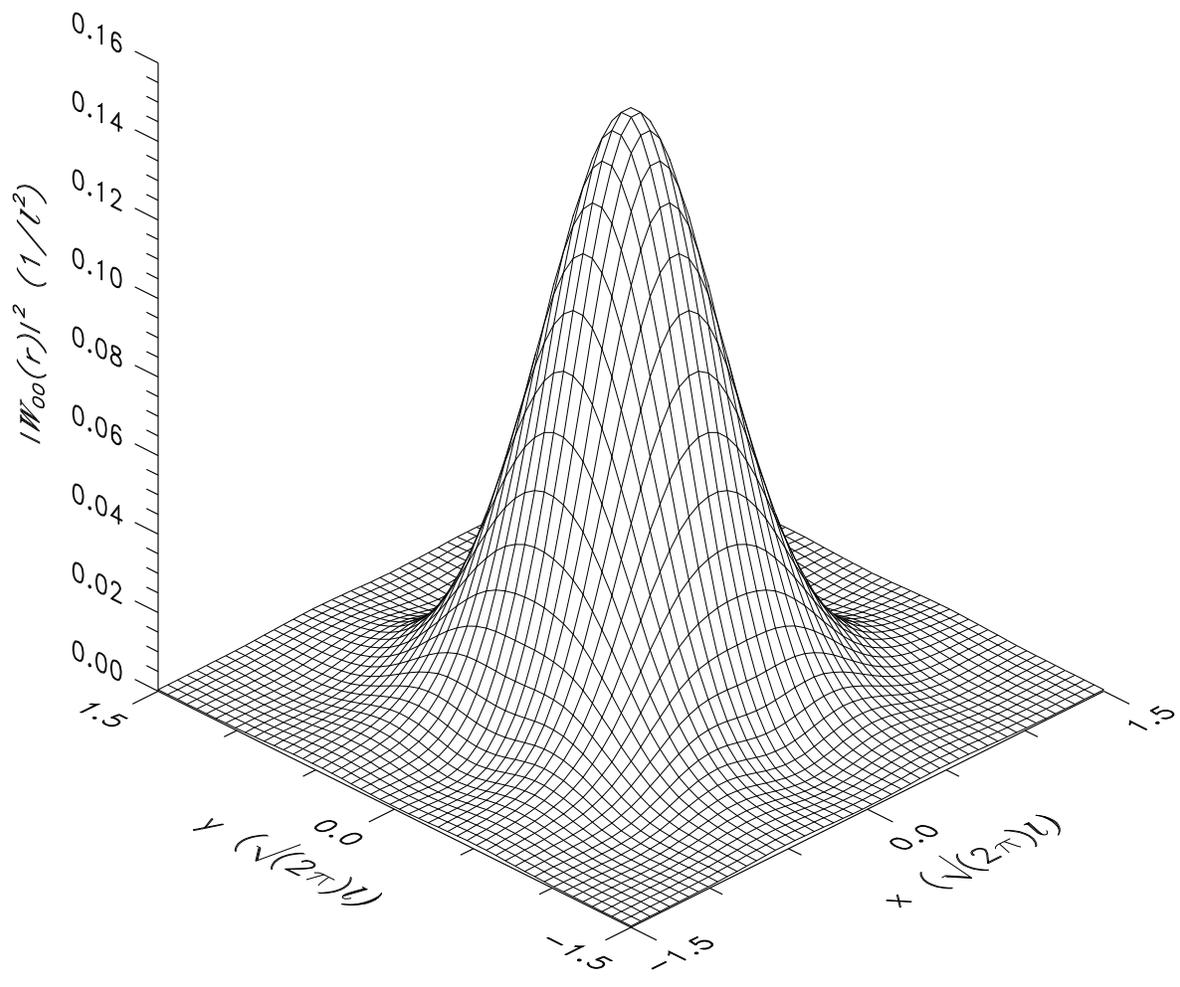
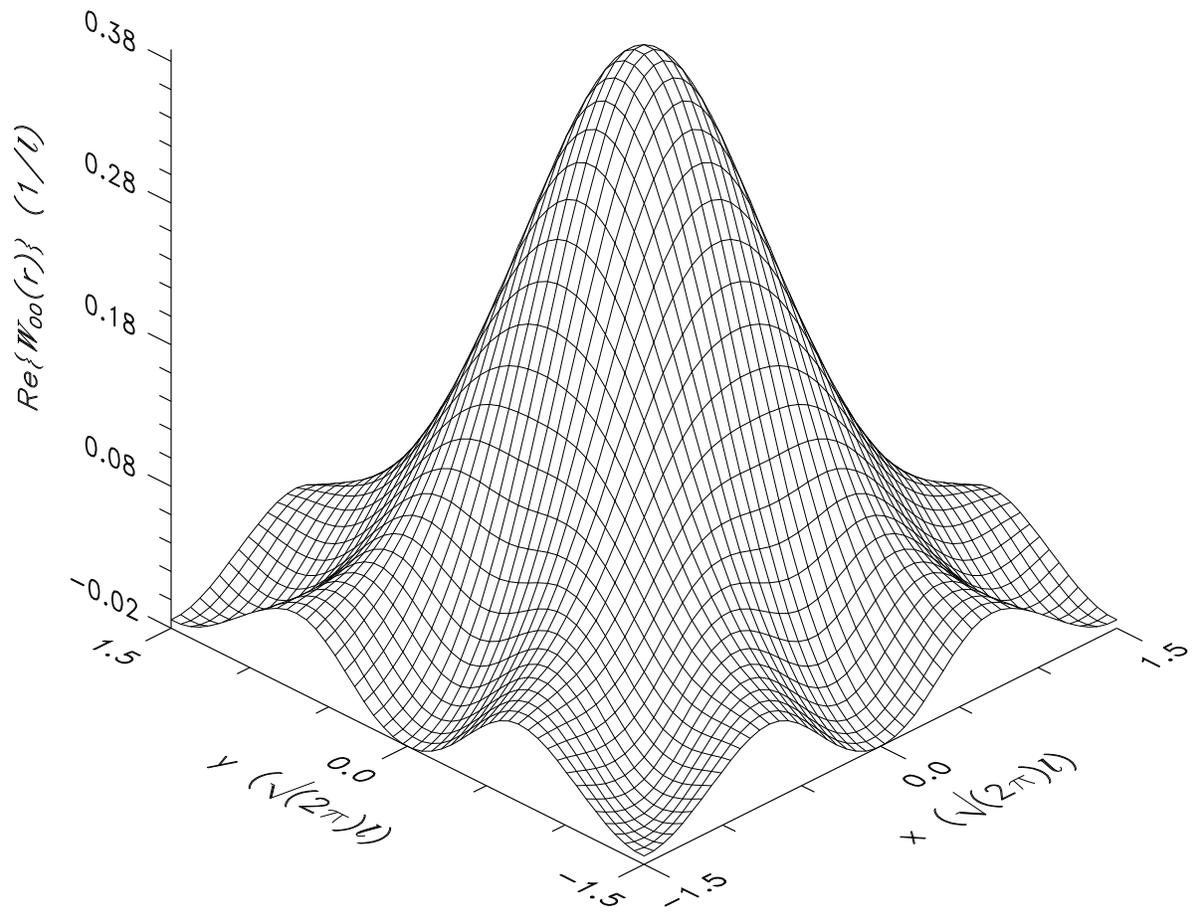
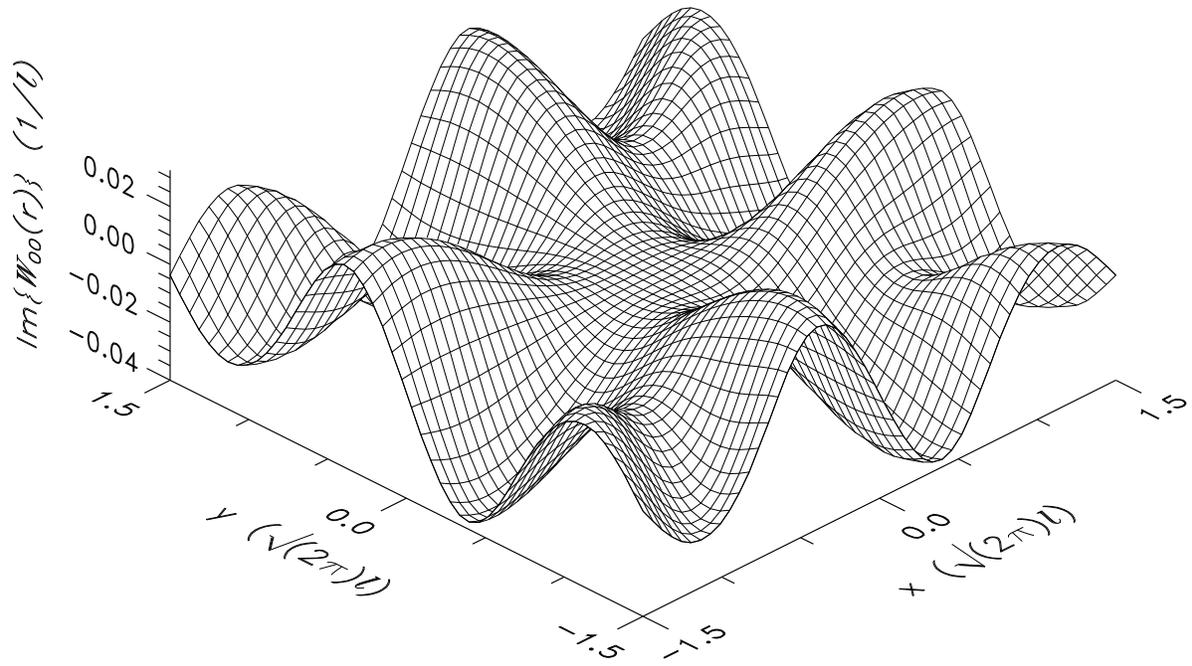
\end{document}